\documentclass[
	a4paper, 
	10pt, 
	twoside, 
]{LTJournalArticle}

\runninghead{} 

\footertext{} 

\usepackage{amsmath,amssymb}
\title{On the Galilean covariance of the d'Alembert equation for acoustic phenomena} 

\author{%
	Francisco Caruso\textsuperscript{1}, Vitor Oguri\textsuperscript{2} and Felipe Silveira\textsuperscript{3}\thanks{Corresponding author: \href{mailto:felipeoak91@gmail.com}{felipeoak91@gmail.com}\\ \textbf{} \textbf{} }
}

\date{\footnotesize\textsuperscript{\textbf{1}}Centro Brasileiro de Pesquisas F\'{\i}sicas -- Rua Dr.~Xavier Sigaud, 150, 22290-180, Urca, Rio de Janeiro, RJ, Brazil.\\ \textsuperscript{\textbf{2}}Instituto de F\'{\i}sica Armando Dias Tavares, Universidade do Estado do Rio de Janeiro -- Rua S\~ao Francisco Xavier, 524, 20550-900, Maracan\~a, Rio de Janeiro, RJ, Brazil.\\ \textsuperscript{\textbf{3}} Independent Researcher.}

\renewcommand{\maketitlehookd}{%
	\begin{abstract}
		\noindent The covariance of the d'Alembert equation for acoustic phenomena -- which is a mechanical wave equation -- under the conventional Galilean transformation is demonstrated without the need to abandon the hypothesis that time is absolute in Classical Mechanics, {what would imply} a modification of Galileo's transformations, as suggested in a paper recently published in this journal.
	\end{abstract}
}

\begin{document}
\maketitle

\section{Introduction}

Until now, the question of the covariance of the d'Alembert equation for acoustic waves has been approached in a wrong way, as will become clear in this paper. Commonly, the subject is addressed in texts on Special Relativity, motivated by the well-known fact that the components of the electric and magnetic vector fields of light satisfy the same d'Alembert equation. As Maxwell's electromagnetic theory is covariant by the Lorentz transformation group~\cite{1, 2}, one may be led to think that the same applies to the d'Alembert equation when it describes any other physical phenomenon. This is indeed the case so far light phenomena are involved. However, {acoustic phenomena are typically non-relativistic}. {Thus one can wonder if there is any contradiction to have both relativistic and non-relativistic physical phenomena described by the same differential equation. How the same equation can sometimes be covariant according to the Lorentz transformations and at other times by the Galileo transformations.} It is important to stress from the beginning that there is no contradiction in the fact that the d'Alembertian equation, applied to acoustic phenomena, is covariant by the transformations of the Galileo group. The physical reason for this remarkable difference {between the relativistic and the non-relativistic limits} is, ultimately, the constancy of light velocity, as proposed by Einstein. {This will become evident in the deduction that follows. Disregarding} this reasoning, this specific issue was recently treated by Berisha and Klinaku~\cite{3}. {The authors} actually claimed to give an answer to the question ``Why does the acoustic wave equation turn out to be non-invariant to Galilean transformations?'', by sustaining that Galileo's covariance of this equation, in the case of acoustic waves, can be accomplished only by defining a new Galilean transformation where time is no more absolute. The statement {rhetorically} put here as a question is absolutely not true, as will be shown in Sec.~\ref{Sec2} and in the sequel.

In general, it is widely known that both the speed of sound and light do not depend on the state of motion of the source. However, unlike light, the speed of sound depends on the movement of the observer. Einstein's hypotheses {that the} speed of light in vacuum {is an universal constante and that the} invariance of physical laws, with respect to transformations between inertial reference frames, condition the covariance of d'Alembert's equation for light waves to Lorentz transformations. Meanwhile, so far acoustic phenomena are considered, the covariance of the d'Alembert equation must be accomplished by the usual Galilean transformations, as it will be demonstrated below.

The Galilean transformation~\cite{4} is a mathematical relationship used in Classical Physics to describe the transformation between the coordinates of an event as observed from two different inertial reference frames. It stems from the so-called Principle of Relativity, first expressed by the Italian scientist Galileo Galilei, during the 17th Century~\cite{5}. In its original conception, it is applicable when the relative velocity between the two frames is much smaller than the speed of light in vacuum, making it appropriate for describing non-relativistic scenarios~\cite{6}.

Without lost of generality, one can consider the Galilean transformation equations for coordinates, considering the relative movement of the reference frames just along the $x$ axis, so that:
\begin{equation}
    \begin{cases}
      x^{\prime} = x - Vt \\
      t^{\prime} = t
    \end{cases}
  \end{equation}

\noindent
where $(x, t$) and $(x^\prime, t^\prime$) are the spatial and temporal coordinates according to the reference frames $S$ and $S^\prime$, respectively, and $V$ is the speed of $S'$ relative to $S$.

The Galilean transformation describes how the spatial and temporal coordinates of an event measured in one reference frame can be related to those measured in the other reference frame. It forms the basis for the Classical Newtonian Mechanics when dealing with non-relativistic speeds, such as those encountered in everyday life situations. However, when dealing with extremely high speeds like those of elementary particles in accelerators, the Lorentz transformation, derived from Einstein's Theory of Special Relativity which defines a four dimensional spacetime, becomes necessary to accurately describe all the changes between two different inertial frames.

Berisha and Klinaku~\cite{3} propose a modified version of the Galilean transformation, which they called the transformed Galilean transformation (TGT), expressed by the {\textit{ad hoc}} relations
\begin{equation}
    \begin{cases}
      x^{\prime} = x - Vt \\
    \displaystyle   t^{\prime} = t - Vx/c^2
    \end{cases}
  \end{equation}

\noindent
where $c$ {here} is sound velocity {in the inertial reference frame $S$, denoted by $u$ in the aforementioned paper}. After deducing the TGT, they use it to demonstrate the invariance of the wave equation for acoustic phenomena.

In this paper, it will be shown that to demonstrate this invariance we don't need to use the TGT.

\section{Galilean covariance of the d'Alembert equation in one dimension}\label{Sec2}

Regarding a coordinate system $S$, associated with a stationary reference frame in a non-dispersive medium, the wave equation governing the acoustic phenomena is given by~\cite{7, 8, 9}

\begin{equation}\label{eq-dalembert}
\displaystyle \frac{1}{c^2} \, \frac{\partial^2}{\partial t^2}  \Psi(x, t) =  \frac{\partial^2}{\partial x^2}  \Psi(x, t)
\end{equation}

\noindent
where $c$ represents the propagation speed of sound, which, in this case, solely depends on the properties of the medium. Here, $(x, t)$ are the spatial and temporal coordinates in reference to $S$, and $\Psi(x, t)$ is a scalar field, such as pressure or density of the medium. Notably, its value at a given instant and point in the medium remains independent of the reference frame.

Consequently, if $(x^\prime, t^\prime)$ represents the space-time coordinates concerning a system $S^\prime$ associated with a non-stationary reference frame, yet with axes parallel to $S$, the following equation must hold, since $\Psi$ is a scalar,
\begin{equation} \label{invar_psi}
 \Psi (x, t) = \Psi (x^\prime, t^\prime)
 \end{equation}

\begin{figure}[tb]
\begin{center}
\includegraphics[width=7.5cm]{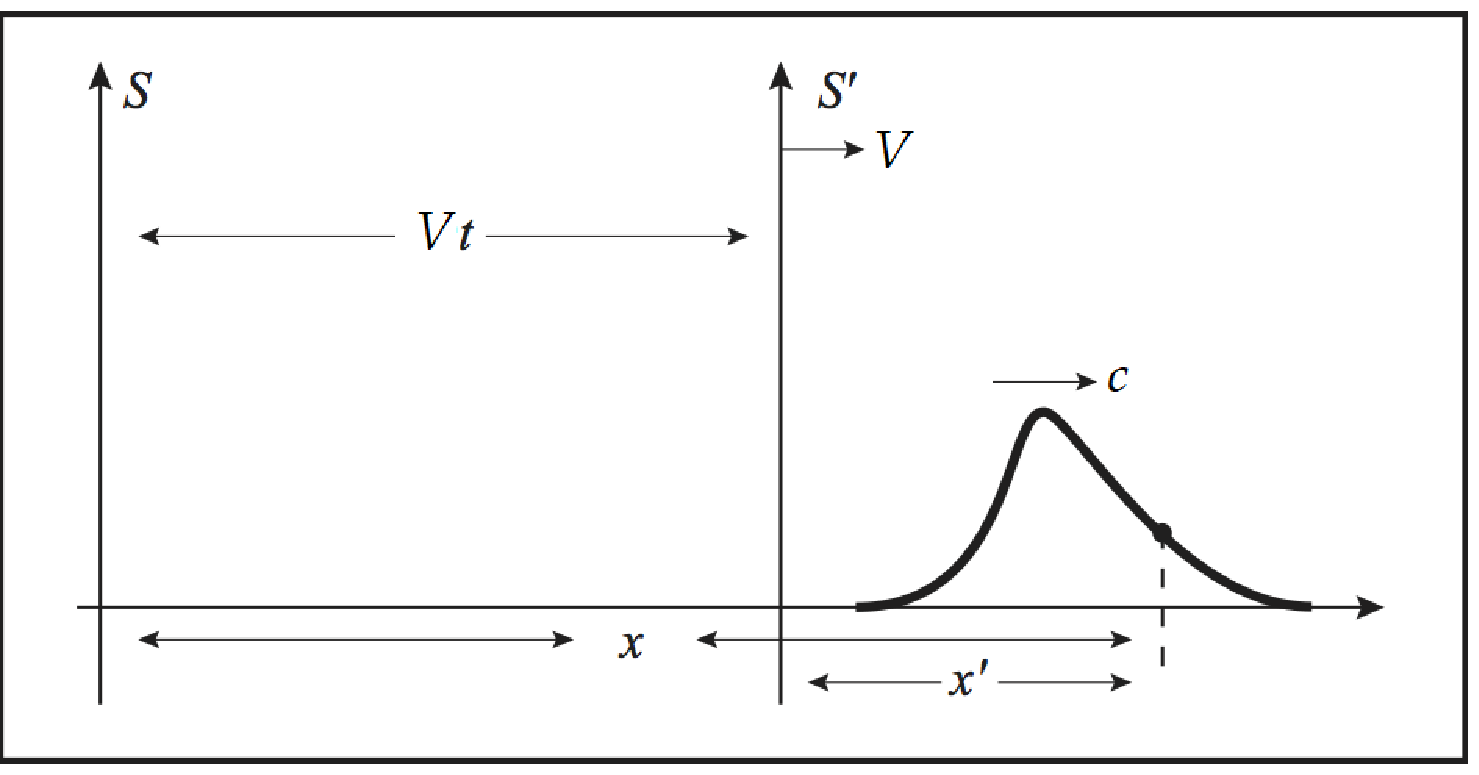}
\caption{The propagation of a sound pulse is observed from two different reference frames. Here, $c$ represents the velocity of the pulse relative to the system $S$, and $V$ is the velocity of $S^\prime$ relative to $S$.}
\label{invar_onda}
\end{center}
\end{figure}

In fact, if the frame associated with $S^\prime$ moves relative to the stationary frame from $t=0$, with a velocity $V$, in the same direction and sense ($+x$) as the propagation velocity ($c$) of a sound pulse (Figure~\ref{invar_onda}), then
$$ \Psi(x, t) = f(x - ct) $$

Furthermore, considering that the clocks in $S$ and $S^\prime$ were synchronized at $t = t^\prime =0$, and remain synchronized such that at any later instant $t^\prime = t$, the coordinates in $S$ and $S^\prime$ are related by

\begin{equation} \label{galileu_t}
x = x^\prime + Vt = x^\prime + V t^\prime
 \end{equation}

According to the hypothesis of invariance of time intervals, the relationship

$$ f(x-ct) = f(x^\prime +Vt^\prime -ct^\prime) =$$ \\ \vspace{-0.8cm} $$= f[x^\prime - (c-V)t^\prime] = f(x^\prime - c^\prime t^\prime) $$
implies the equation~(\ref{invar_onda}), where $c^\prime = (c - V)$ is the speed of sound relative to the system $S^\prime$, \textit{i.e.}, the speed of sound with respect to a reference frame that moves in a medium, in addition to the properties of the medium, also depends on the speed of the reference frame.

Regarding the wave equation~(\ref{eq-dalembert}), from the relationships between the spatial and temporal derivatives in the $S$ and $S^\prime$ systems, one should write

$$ \left\{
\begin{array}{l}
\displaystyle
\frac{\partial}{\partial x^\prime} = \overbrace{\frac{\partial x}{\partial x^\prime}}^{1} \frac{\partial}{\partial x}
\quad \Rightarrow \quad \frac{\partial^2}{\partial {x^\prime}^2} = \frac{\partial^2}{\partial x^2} \\
 \ \\
\displaystyle
\frac{\partial}{\partial t^\prime} = \overbrace{\frac{\partial t}{\partial t^\prime}}^{1} \frac{\partial}{\partial t} \ + \ \overbrace{\frac{\partial x}{\partial t^\prime}}^{V} \frac{\partial}{\partial x} \\
\ \\
\displaystyle
\Rightarrow \quad
 \frac{\partial^2}{\partial {t^\prime}^2} = \frac{\partial^2}{\partial t^2} + 2 V \frac{\partial^2}{\partial t \partial x} + V^2 \frac{\partial^2}{\partial x^2}
  \end{array}
\right.
$$
As a result:
$$ \left\{
\begin{array}{l}
\displaystyle
\frac{\partial^2}{\partial {x^\prime}^2} \Psi(x^\prime, t^\prime) = \frac{\partial^2}{\partial x^2} \Psi(x, t)  \\
 \ \\
\displaystyle
\frac{\partial^2}{\partial {t^\prime}^2} \Psi(x^\prime, t^\prime) =  \frac{\partial^2}{\partial t^2} \Psi(x, t) \ + \\
\ \\
\displaystyle
\hspace*{1.2cm} + \ 2 V \frac{\partial^2}{\partial t \partial x} \Psi(x, t) + V^2 \frac{\partial^2}{\partial x^2} \Psi(x, t)
  \end{array}
\right.
$$

Taking into account that
$$ \Psi(x, t) = f(x - ct) \quad \Rightarrow \quad
\left\{
\begin{array}{l}
\displaystyle
 \frac{\partial^2}{\partial t^2} \Psi(x, t) = c^2  f^{\prime \prime} \\
 \ \\
\displaystyle
 \frac{\partial^2}{\partial t \partial x} \Psi(x, t) = - c  f^{\prime \prime}   \\
 \ \\
\displaystyle
\frac{\partial^2}{\partial x^2} \Psi (x, t) = f^{\prime \prime}
  \end{array}
\right.
$$
finally yielding
$$
\displaystyle
\frac{\partial^2}{\partial {t^\prime}^2} \Psi(x^\prime, t^\prime) = \big( \underbrace{c^2 - 2cV + V^2}_{(c-V)^2} \big) f^{\prime \prime} = {c^\prime}^2 \, \frac{\partial^2}{\partial {x^\prime}^2} \Psi(x^\prime, t^\prime)
$$
being $c^\prime = c - V$.

Thus, the sound wave propagation equation has the same form in any inertial reference frame. Alternatively, this property is expressed by saying that the d'Alembert equation for acoustic waves is covariant with respect to the Galilean transformations.

$$
\fbox{~$\displaystyle \frac{1}{{c^\prime}^2} \, \frac{\partial^2}{\partial {t^\prime}^2} \Psi(x^\prime, t^\prime) =
 \frac{\partial^2}{\partial {x^\prime}^2}  \Psi(x^\prime, t^\prime)$~}
$$

\vspace*{0.5cm}

\section{Phase Invariance and Doppler Effect}

In the case of a monochromatic sound wave of wavelength $\lambda = 2 \pi/k$ and frequency $\omega=2\pi \nu = k c$, described by
$$\Psi(x, t) = \cos \frac{2 \pi}{\lambda} ( x- ct) = \cos (k x - \omega t)$$

\noindent
the validity of the expression $ f(x-ct) = f(x^\prime - c^\prime t^\prime)$ implies that the phase of the wave function is invariant~\cite{10} with respect to changes in space-time coordinates $(x, t)$ e $(x^\prime, t^\prime)$ in the $S$ and $S^\prime$ systems. So, according to equation~(\ref{galileu_t}),
 $$ k^\prime x^\prime - \omega^\prime t^\prime = k x - \omega t = k x^\prime + k V t^\prime - \omega t^\prime =$$ \\ \vspace{-0.9cm} $$= k x^\prime -(\omega - kV) t^\prime  $$
 one gets
 $$
\left\{
\begin{array}{l}
\displaystyle
 k^\prime = k \quad \Rightarrow \quad  \lambda^\prime = \lambda \\
 \ \\
\displaystyle
\omega^\prime =  \omega - k V = \omega \left( 1 - \frac{V}{c} \right)
  \end{array}
\right.
$$

\noindent as it should be. Remember that the wave-length, by definition, is the {spatial} distance between two successive crests of a wave, which is invariant under the original Galilean transformation. In this way, the relation between the wave propagation speed, wave-length and frequency is {also} invariant. Indeed,
$$ c^\prime = \lambda^\prime \nu^\prime = (c-V) = \lambda \nu \left(1 -\frac{V}{c} \right) \quad \Rightarrow \quad c = \lambda \nu$$

The dependence of the frequency on the reference frame, called the Doppler Effect\footnote{In this case, for the source at rest in $S$, and an observer \\ moving in the same direction as the wave.}~\cite{7, 8}, is a phenomenon characteristic of the propagation of any type of wave, be it acoustic or electromagnetic. For the latter, $c$ is a universal constant, and in the case of the source or the observer approaching with velocity $V$, the wave-length changes like any other distance along the direction of motion, $\lambda^\prime = \lambda \sqrt{(1 - V/c)/(1 + V/c)}$, and the frequency goes to $\nu^\prime = \nu \sqrt{(1 + V/c)/(1 - V/c)}$, such that the relation $ c = \lambda \nu$ is preserved in any inertial frame. Thus, no matter the nature of the phenomenon described by the d'Alembert equation, the invariance of the relation $c^\prime = \lambda^\prime \nu^\prime = \lambda \nu = c$ is always verified.

In the case of sound waves, this dependence is a direct consequence of the relative motion between the observer and the sound source.

For light waves in a vacuum, even though the speed of light does not depend on the relative motion between the observer and the light source, the effect manifests itself due to the relativity of time intervals in different reference frames.

\section{Discussion}

{It was straightforwardly demonstrated} that, {in order to show the invariance of the wave equation for acoustic phenomena,} there is no need to modify the Galilean transformation, contrary to what is {sustained} in Berisha and Klinaku work~\cite{3}. This finding indicates that, {under the classical Galilean transformation,} the form of the d'Alembert equation remains unaltered in different inertial reference frames moving at constant velocities relative to each other, provided that those velocities are much smaller than the speed of light, \textit{i.e.}, {in} non-relativistic scenarios).

\end{document}